\documentclass[useAMS,usenatbib]{mn2e} %
\IfFileExists{srcltx.sty}{\usepackage[active]{srcltx}}

\usepackage{amssymb}% 
\usepackage{amsmath}% 
\usepackage[dvips]{graphicx}% 
\usepackage{natbib} %
\usepackage{url,varioref,lastpage,bm,color}

\newif\ifpdf\ifx\pdfoutput\undefined\pdffalse\else\pdfoutput=1\pdftrue\fi

\usepackage[dvips,colorlinks=true]{hyperref}

\let\jnlstyle=\rm
\def\jref#1{{\jnlstyle#1}}

\def\apj{\jref{ApJ}}                 
\def\apjl{\jref{ApJ}}                
\def\apjs{\jref{ApJS}}

\def\aap{\jref{A\&A}}                
          
\def\aaps{\jref{A\&AS}}

\def\mnras{\jref{MNRAS}}

\newcommand{\fov}{{\mathrm{fov}}} % Field-of-view
\newcommand{\cl}{{\mathrm{cluster}}} % Cluster
\newcommand{\vir}{{\mathrm{vir}}} % Virial
\newcommand{\dm}{{\textsc{dm}}} % Dark Matter
\newcommand{\kev}{\:\mathrm{keV}} % keV
\newcommand{\mpc}{\:\mathrm{Mpc}} % Mpc
\newcommand{\const}{\mathrm{const}} %

\begin{document}
\title{Restrictions on parameters of sterile neutrino dark matter from
  observations of galaxy clusters.}  %
\author[Boyarsky et
al.]{A.~Boyarsky$^{1,2,3}$, A.~Neronov$^{4,5}$, O.~Ruchayskiy$^{6}$,
  M.~Shaposhnikov$^{2,1}$
  \\
  $^{1}$CERN, Theory department, Ch-1211 Geneve 23,
  Switzerland\\
  $^{2}$\'Ecole Polytechnique F\'ed\'erale de Lausanne, Institute of
  Theoretical Physics,\\~~FSB/ITP/LPPC, BSP 720, CH-1015, Lausanne,
  Switzerland\\
  $^{3}$On leave of absence from Bogolyubov Institute of Theoretical Physics,
  Kyiv, Ukraine\\
  $^{4}$INTEGRAL Science Data Center, Chemin d'\'Ecogia 16,
  1290 Versoix, Switzerland\\
  $^{5}$Geneva Observatory, 51 ch. des Maillettes,
  CH-1290 Sauverny, Switzerland \\
  $^{6}$Institut des
  Hautes \'Etudes Scientifiques, Bures-sur-Yvette, F-91440, France\\
}

\date{Received $<$date$>$ ; in original form $<$date$>$ }
\pagerange{\pageref{firstpage}--\pageref{LastPage}} \pubyear{2005}

\maketitle
\label{firstpage}

\begin{abstract}
  We find restrictions on the mass and mixing angle of the dark matter
  sterile neutrinos using X-ray observations of Coma and Virgo galaxy
  clusters.  
  %Contrary to the conclusions of the previous works we find that
  %this bound is not stronger than the one obtained from the analysis of the
  %X-ray background by \cite{Boyarsky:05} and explain the reason for that.
%  The reason for this is that the emission from the cores X-ray bright
%  clusters is dominated by the thermal bremsstrahlung continuum and line
%  emission from the hot intracluster medium. Based on this observation we find
%  that e.g. for Virgo cluster a slightly better bound on sterile neutrino
%  parameters can be obtained from the analysis of the region outside the core
%  of X-ray emission.
\end{abstract}

\section{Introduction}
\label{sec:decay}

Recent experiments on neutrino oscillations (for a review see
e.g.~\cite{Strumia:05}) determine non-zero difference of masses of
various species of neutrinos. The experiments on atmospheric
neutrinos~\citep{superK} are explained via $\nu_\mu\to\nu_\tau$
oscillations with mass difference $\Delta m^2_{\mathrm{atm}} =
[2.2^{+0.6}_{-0.4}]\cdot 10^{-3}\:\mathrm{eV}^2$, while
experiments~\citep{SNO,kamland} determine mass squared difference
$\Delta m^2_{\mathrm{sol}} = [8.2^{+0.3}_{-0.3}]\cdot
10^{-5}\:\mathrm{eV}^2$ for $\nu_e\to\nu_{\mu,\tau}$ transitions. 
Probably, a simplest way to consistently explain these data is to
add  to the Standard Model several (at least two) gauge singlet
fermions - right handed, or sterile neutrinos. However, the absolute
values of masses of active and sterile neutrinos are not fixed by
these experiments.

It was noticed long time ago (c.f.~\cite{Dodelson:93}) that a sterile
neutrino with the mass in the keV range appears to be a viable
``warm'' DM candidate. If such a neutrino is a main ingredient of the
DM, it is potentially detectable in various X-ray observations. To
wit, there exists a (sub-dominant) radiative decay channel of sterile
neutrino $N$ into active neutrinos $\nu_a$ and photon with energy
$E=m_s/2$, with the width~\citep{Pal:81,Barger:95}
\begin{equation}
\label{gamma}
  \Gamma_{N\to\gamma\nu_a} = \frac{9\, \alpha\, G_F^2}
  {256\cdot 4\pi^4}\sin^22\theta\,
  m_s^5 = 
  5.5\times10^{-22}\theta^2
  \left[\frac{m_s}{1\,\mathrm{keV}}\right]^5\;\mathrm{s}^{-1}\:.
\end{equation}
Here $\theta$ is the mixing angle defined as 
\begin{equation}
\theta^2=\frac{1}{m_s^2}\sum_{\alpha=\{e\mu\tau\}}|M_D|_{s\alpha}^2
\end{equation}
with $M_D$ being the Dirac mass and $m_s$ is the sterile neutrino mass.

Although even for the main decay channel ($N\to 3\nu_a$) the lifetime of
sterile neutrino must exceed the age of the universe, the decay $N\to
\nu_a\gamma$ can result in a potentially detectable X-ray flux
\citep{Dolgov:00,Abazajian:01b}, since the density of DM in the universe is
large. For example, the contribution to the diffuse X-ray background (XRB) can
be comparable with measured XRB and is given by (see~\cite{Boyarsky:05} and
references therein)
\begin{equation}
F_{\rm XRB}\simeq \frac{\Gamma \rho_{\dm}^0}{2\pi H_0}\simeq
5\times 10^{-4}\,\theta^2\left[\frac{m_s}{1\:
      \mathrm{keV}}\right]^5\frac{\mbox{erg}}{\mathrm{cm^2\cdot s\cdot
      sr}}~, 
\end{equation}
where $\rho^0_{\dm}, H_0$ are the dark matter density in the universe and
the Hubble constant. Neutrinos decaying at different red shifts produce a broad
X-ray line with extended ``red'' tail.  Such a feature in the XRB spectrum is,
in principle, readily detectable (and distinguishable) from the broad-band
continuum of observed XRB~\citep{Gruber:99}. The non-detection of the DM decay
feature in the XRB signal enables to put a bound on $\theta$ and $m_s$ roughly
at the level of~\citep{Boyarsky:05}
\begin{equation}
\label{xrbbound}
\Omega_s \sin^2(2\theta)\lesssim 3\times 10^{-5}
\left[\frac{1\mbox{ keV}}{m_s}\right]^{5}~,
\end{equation}
where $\Omega_s$ is the present day density of sterile neutrino, understood as
DM candidates.

Clustering of the DM at small red shifts results in the enhancement of the DM
decay signal in the direction of large mass concentrations, such as galaxy
clusters. Taking into account the typical overdensity ${\cal R}=\rho/\rho_{\rm
  DM}^0$ of a galaxy cluster is at the level of ${\cal R}\sim 10^3$, while
typical cluster size is about $D\sim$~Mpc~$\sim 10^{-3}H_0^{-1}$, one can find
that the DM decay flux from a galaxy cluster,
\begin{equation}
  \label{eq:15}
  F_\cl =\frac{M_\dm^\fov\Gamma}{8\pi D_L^2}
\end{equation}
($M_\dm^\fov\simeq {\cal R}\rho^0_{\dm}DD_\theta^2\Omega^\fov$ is the mass
of DM within telescope's field of view (FoV) and $D_L,D_\theta$ are the
luminosity and angular diameter distances to the cluster) is comparable to the
background DM decay signal
\begin{equation}
\frac{F_\cl}{F_{\rm XRB}}\sim {\cal R}DH_0\sim 1\;.
\end{equation}%$
Although the total DM decay flux from the XRB and from the cluster are of the
same order, their spectra are different. The flux from the cluster would be
detected as a narrow line whose width $\Delta E=\Delta E_{\rm det}$ is
determined by the spectral resolution of an X-ray detector. At the same time,
the DM decay contribution into XRB is produced by the decays at red shifts
$z\sim 0\div 1$ and, as a result the DM decay line is broadened to $\Delta
E\sim m_s/2$. Thus, in spite of the fact that the compact DM sources at
$z\simeq 0$ give just moderate enhancement of the DM decay flux, the
enhancement of the signal in the narrow energy band centered on the line
energy $E=m_s/2$ could be large. From the energy dependence of $dF_{XRB}/dE$
(see e.g.~\cite{Boyarsky:05}) one can estimate that the enhancement can be by
a factor of $m_s/\Delta E_{\rm det}\sim 10-100$ for instruments of spectral
resolution of Chandra or XMM (in this paper we will use XMM observations
of Virgo and Coma clusters).

However, the detection of the DM decay line in the galaxy clusters is
complicated by the fact that most of the galaxy clusters show strong
continuum and line emission from the hot intracluster gas. The typical
temperatures of the intraculser gas,
\begin{equation}
T_{\rm gas}\sim G\ {\cal R}\rho^0_{\dm}D^2m_p\sim
10\left[\frac{{\cal R}}{10^3}\right]\left[\frac{D}{1\mbox{
      Mpc}}\right]^2\mbox{ keV}\;,
\end{equation} 
lie exactly the range of \emph{Chandra} or XMM-Newton satellites -- 1-10 keV.
In fact, for a typical nearby X-ray bright cluster, the diffuse continuum
emission from the cluster core is by a factor $\sim 10^2-10^3$ stronger than
the XRB emission from behind the core. Thus, the factor $10-100$ increase of
the flux in the DM decay line from the cluster direction is ``washed out'' by
the strong increase of the X-ray continuum emission from the direction of a
cluster.  We will see however than in spite of such a strong background, the
bound on $\sin^22\theta$ can be improved by a factor of about 2--4 in
comparison with XRB~(\ref{xrbbound}).

%Such simple qualitative argument enables to conclude that the bound
%on the sterile neutrino parameters which would come from the observations of
%the cores of X-ray bright clusters would not be much better then the bound
%(\ref{xrbbound}) obtained from the analysis of the extragalactic X-ray
%background.

Previously this question was addressed in the paper by
\cite{Abazajian:01b,Abazajian:05a}.  There the authors claimed the following
limit on the neutrino mass and mixing angle, derived from the non-observation
of the DM decay line from the center of the Virgo galaxy cluster:
$\sin^2(2\theta) \lesssim 1.0\times 10^{-5} (\kev/m_s)^4$ for $1< m_s <
10$~keV. This limit is several times better than Eq.~(\ref{xrbbound}).

In this paper we re-analyze the bounds on the sterile neutrino mass and mixing
angle imposed by the observations of Coma and Virgo galaxy clusters.  As we
will discuss in Section~\ref{sec:results}, our constraints are weaker than
those claimed by~\cite{Abazajian:01b,Abazajian:05a}. We will compare in
details these two results in Section~\ref{sec:concl} and explain the origin of
the difference.

In this work we show that the problem of the large continuum X-ray emission
from the cluster core can be partially relaxed, if one searches for the DM
decay signal from a region outside the bright X-ray core of the cluster. The
reason for that is the following: outside the core of the cluster, the surface
brightness profile of intracluster medium (ICM) behaves as ${\cal S}(r) \sim
r^{-\alpha}$ (where, e.g. for Virgo cluster $\alpha \sim
1.6$~\citep{Fabricant:83,Schindler:99,Young:02}), while the surface brightness
profile of the DM is more shallow ${\cal S}_\dm(r) \sim r^{-1}$~(see
e.g.~\citet{Cavaliere:76}, for mass density of Virgo cluster).  Therefore, by
moving away from the center, one can put much tighter constraints on the DM
parameters. For the case of Coma cluster observations with XMM one needs to
use periphery observations (``coma mosaic'' observations, performed in
May-June 2000, \citep{Briel:00,Neumann:02}).  For the case of Virgo this goal
can be achieved already for the XMM observation of M87 (observation ID
0114120101) Indeed, as the core radius of Virgo cluster is very small ($r_0
\lesssim 1.6'$), the existing observations with XMM EPN camera (field-of-view
$\sim 15'$) allows one to study properties of Virgo cluster till about $10
r_0$.  We also show that the bound can be further improved if the X-ray
spectra of galaxy clusters are analyzed by the method proposed by
\citet{Boyarsky:05} for the search of the DM decay line in the XRB spectrum.
We describe the analysis of the data from Coma and Virgo observations in
Section~\ref{sec:analysis} and present results in Section~\ref{sec:results}.

%%%%%%%%%%%%%%%%%%%%%%%%%%%%%%%%%%%%%%%%%%%%%%%%%%%%%%%%%%%%%%%%%%%%%
\section{DM decay line vs. thermal bremsstrahlung emission.}
\label{sec:theory}
%%%%%%%%%%%%%%%%%%%%%%%%%%%%%%%%%%%%%%%%%%%%%%%%%%%%%%%%%%%%%%%%%%%%

The main obstacle for improving the bound on the neutrino parameters from the
X-ray background by using 
galaxy clusters observations is strong X-ray continuum and line emission from
the cluster core. In this Section we show that this obstacle could be
partially avoided if one extracts the X-ray spectrum from the regions beyond
the core of the continuum X-ray emission. The idea is that the radial surface
brightness profile of the continuum emission is
normally steeper than the radial surface brightness profile expected from the
DM decay line.

Indeed, the surface brightness profile of a relaxed cluster of galaxies is
usually well-fitted by the so-called isothermal
$\beta$-model~\citep{Cavaliere:76,Sarazin:77}:
\begin{equation}
  \label{eq:1}
  {\cal S}(r)=\frac{{\cal S}_0}{(1+(r/r_0)^2)^{3\beta-0.5}}\;,
\end{equation}
where $r$ is the \emph{projected} distance from the center of the cluster,
${\cal S}_0$ is the overall normalization factor and $r_0$ is the ``core
radius'' of the region of X-ray emission. Parameters $\mathcal{S}_0$, $r_0$ and
$\beta$ are different for different clusters (see Section~\ref{sec:analysis}
below).

The volume emissivity of the intracluster gas which results in the surface brightness profile (\ref{eq:1}) is
\begin{equation}
  \label{volume}
  {\cal V}(\bm{r})=\frac{{\cal V}_0}{(1+(\bm{r}/r_0)^2)^{3\beta}}~,
\end{equation}
where $\bm{r}$ is the (3d) distance to the center of the cluster.
The continuum X-ray emission from the intracluster gas is the thermal
bremsstrahlung for which the volume emissivity is proportional to the square
of the gas density. The radial density profile of the hot intracluster gas is
therefore
\begin{equation}
  \label{eq:3}
  n_\mathrm{gas}(\bm r)=\frac{n_0}{(1+(\bm r/r_0)^2)^{3\beta/2}}\sim
  \sqrt{{\cal V}(\bm r)}~.
\end{equation}
To calculate the radial mass profile $M(r)$ (gas plus galaxies plus the DM) of
the cluster one has to make an assumption that the intracluster gas is in
hydrostatic equilibrium~\citep{Cavaliere:76,Sarazin:77}. One can really expect
this assumption to hold only for relaxed galaxy clusters, but as it turns out
even for e.g. Virgo this model gives good predictions (c.f.~\cite{Nulsen:95}).
Under this assumption the overall mass profile can be calculated from the
Newton's law
\begin{equation}
  \label{eq:4}
  \frac{dp}{d\bm{r}}=n_\mathrm{gas}(\bm{r})\frac{dT(\bm{r})}{d\bm{r}}+T(\bm{r})\frac{dn_\mathrm{gas}(\bm{r})}{d\bm{r}}=
  -\frac{GM(\bm{r})n_\mathrm{gas}(\bm{r})}{\bm{r}^2}~,   
\end{equation}
where $T(\bm r)$ is the radial temperature profile of the hot intracluster
gas.  In the region where the temperature does not change significantly, one
can calculate the overall mass profile analytically by
substituting~(\ref{eq:3}) into~(\ref{eq:4}).  The assumption of hydrostatic
equilibrium them immediately leads to the following total mass dependence on
the distance $\bm{r}$ from the center of the
cluster~\citep{Cavaliere:76,Sarazin:77}
\begin{equation}
  \label{eq:2}
  M(\bm{r}) = 1.13\times
  10^{14}M_\odot\:\beta\frac{T}{1\:\kev}\frac{\bm{r}}{\mpc}\frac{( \bm
    r/r_0)^2}{1+(\bm r/r_0)^2}~.
\end{equation}
%\begin{equation}
%  \label{eq:5}
%  M(r)=M_0\frac{r^3}{r_0^3\left(1+(r/r_0)^2\right)}
%\end{equation}
The overall density profile is given by
\begin{equation}
  \label{eq:6}
  \rho_{\rm total}(\bm r)=\rho_{0,\rm total}\frac{3+(\bm r/r_0)^2}
{\Bigl(1+(\bm r/r_0)^2\Bigr)^2}~.
\end{equation}
In the first approximation one can estimate the DM density as $\rho_\dm =
\kappa \rho_{\mathrm{total}}$ with $\kappa\lesssim 1$.

The surface brightness profile~(\ref{eq:1}) should be compared with the
surface brightness of the DM.  Integrating the volume emissivity of the DM
decay line over the line of sight one can find the brightness profile of the
DM decay line
\begin{equation}
  \label{eq:7}
  {\cal S}_\dm(r)=\frac{\Gamma m_s {\cal N}_\dm(r)}{8\pi D_L^2}~,
\end{equation}
where ${\cal N}_\dm(r)=2\int_0^\infty dz\,n_{\dm}(\sqrt{r^2+z^2})$ is the
column density of the DM as a function of the distance from the cluster
center. Substituting $\rho_{\rm total}$ from (\ref{eq:6}) one obtains
\begin{equation}
  \label{eq:8}
 {\cal S}_\dm(r) = \frac{\Gamma\rho_{0,\dm}r_0 }{ D_L^2} 
  \left[\frac{2+(r/r_0)^2}{8(1+(r/r_0)^2)^{3/2}}\right] ~.
\end{equation}
Integrating $2\pi r\mathcal{S}_\dm(r)$~(\ref{eq:8}) from $r=0$ to some $r$ one
finds the flux of DM from a FoV circle with the projected radius~$r$:
\begin{equation}
  \label{eq:10}
  F_{\dm}(r) = \frac\pi4\frac{ \rho_{0,\dm} r_0^3
  \Gamma}{D_L^2}\,g(r/r_0)~,
\end{equation}
where we have defined a \emph{geometry factor} $g(r)$, describing that part of
the cluster, which we are using for the measurement. In case of the circular
region with the projected radius $r$ one gets from~(\ref{eq:8}):
\begin{equation}
  \label{eq:11}
  g(r/r_0) = \frac{(r/r_0)^2}{\sqrt{1+(r/r_0)^2}} ~.
\end{equation}
One can see that ${\cal S}_{\dm}(r)\sim r^{-1}$ for $r\gg r_0$. Thus, the
radial profile of DM decay line is significantly more shallow than the profile
of the continuum X-ray emission from the hot intracluster gas:
\begin{equation}
\label{sb}
\frac{{\cal S}_\dm}{{\cal S}_{\rm gas}}=
\frac{{\cal S}_\dm(0)}{{\cal S}_{\rm gas}(0)}\left[2+\frac{r^2}{r_0^2}\right]\left[
1+\frac{r^2}{r_0^2}\right]^{3\beta-2}\hskip -2ex 
\sim \left(\frac{r}{r_0}\right)^{6\beta-2}\hskip -2ex ,\,r\gg r_0 ~.
\end{equation}
As it is clear from Eq.~(\ref{eq:1}), $\beta < \frac16$, therefore, the share
of DM contribution to the total brightness (at energy $E=m_s/2$) increases as
one moves away from the center of a cluster.

\section{Data Analysis}
\label{sec:analysis}

In this work we have chosen to analyze data of XMM-Newton observations of
Virgo galaxy cluster (observation ID 0114120101, June 2000) and Coma cluster.
For the latter we have taken two observations which constituted the part of
the so called \emph{Coma mosaic}~\citep{Briel:00,Neumann:02}: that of the center of Coma
cluster (observation ID 0124711401, May 2000) and one of the peripheral
observations (``Coma 3'', ID 0124710301, June 2000). 

\subsection{``Total flux'' restrictions.}
\label{sec:total}

First, we find the restrictions on the parameters of sterile neutrino, based
on the total emission from the ICM.  Namely, using the XMM observation data
and estimating a mass of DM in the center of a cluster as in
Section~\ref{sec:theory}, we restrict parameters $m_s$ and $\sin^2(2\theta)$
by demanding that the flux of DM decay line~(\ref{eq:10}) did not exceed the
total flux in the energy bin, equal to the $3\sigma$ where with of the
Gaussian line $\sigma$ is equal to the spectral resolution of XMM. We will
often dub this method simply \emph{total flux restrictions}. This method
provides the
most robust exclusion.

\subsubsection{Analysis for Coma cluster}
\label{sec:total-coma}

\begin{figure}
  \centering
  \includegraphics[width=\linewidth]{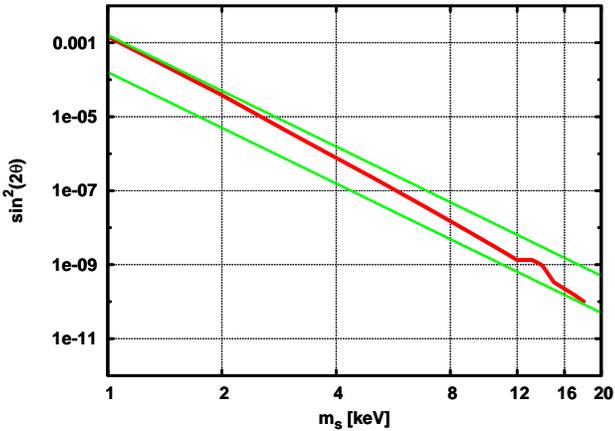}%
  \caption{Exclusion plot from the total flux restriction  from the circular
    region with radius $700''$ in the center of Coma cluster. The straight
    green lines represent $\sin^2(2\theta) m_s^5 = \const$ lines,
    corresponding to
    the total flux at energies 1~keV and 10~keV (in the 200 eV energy bin).}%
  \label{fig:coma-center-total-flux}
\end{figure}

To find the typical numerical values of the DM decay flux as compared to the
thermal bremsstrahlung flux, let us use the parameter values for the Coma
galaxy cluster. Its surface brightness profile is well defined by the
$\beta$-model~(\ref{eq:1}) with $\beta=0.75$ and $r_0 = 10.5' =
0.3$~Mpc\footnote{We take $h=0.71$ and the distance to the Coma cluster
  $D_L=98$~Mpc (red-shift $z=0.023$).} (see e.g.  \cite{Briel:92,Neumann:02}).
The virial mass of Coma cluster, contained in the virial radius $r^\vir_{Coma}
= 3.5$~Mpc is $M^\vir_{Coma}=1.3\times
10^{15}\;\mathrm{M}_\odot$~\citep{Briel:92}.  Substituting these numbers into
Eqs.~(\ref{eq:2})--(\ref{eq:6}) and taking fraction of gas in the Coma cluster
18\%~\citep{Briel:92} one finds
that  %\footnote{ As an example we will
%  analyze below the X-ray observations of Coma cluster of galaxies for which
%  the corresponding parameters are found to be~ $\beta=0.75\pm 0.03$ and $r_0=10.5\pm
%  0.6$~arcmin which is $(0.42\pm 0.024)h_{50}^{-1}=(0.296\pm0.017)$~Mpc.}$^,$
\begin{equation}
\rho_{0,\dm}
\simeq 3\times 10^{14}\;\mathrm{\frac{M_\odot}{Mpc^3}}~.
\end{equation}
Substituting these values in Eq.~(\ref{eq:10}) one finds that the flux of DM
from a FoV circular region with projected radius $r$ is given by
\begin{equation}
\label{DMcoma}
F_{\dm, Coma}\simeq 6.7\times 10^{-8}\,\theta^2\left[\frac{m_s}{1\mbox{
      keV}}\right]^5g(r) 
      \frac{\mbox{erg}}{\mbox{ cm}^2\cdot\mbox{s}} ~,
\end{equation}
where $g(r)$ is given by Eq.(\ref{eq:11}).  Flux~(\ref{DMcoma}) should be
compared with the emission of the ICM in the Coma center region. We use
observation of Coma center with EPN camera of XMM-Newton satellite
(observational ID 0124711401) to extract flux per energy bin, equal to the
spectral resolution of XMM ($\Delta E_{\mathrm{det}} \sim 200$~eV) in the
interval 0.5-10~keV. The geometry factor for the circle of $700''$ is equal to
$0.82$. The resulting restriction on $m_s$ and $\sin^2(2\theta)$ is shown on
Fig.~\ref{fig:coma-center-total-flux} in the red solid line.  Qualitatively
this result can be described as follows. The flux (per energy bin) from Coma
center\footnote{Compare \cite{Briel:92}: $F_{1\kev} \simeq 1.1\times
  10^{-11}\text{erg}/\text{cm}^2\cdot\text{s}$ in 200~eV energy bin.}  at
1~keV $F_{1\kev}=1.62\times 10^{-11}$~erg/cm$^2\cdot$s is about an order of
magnitude bigger, than that at 10~keV $F_{10\kev} = 2.5\times
10^{-12}$~erg/cm$^2\cdot\mathrm{s}$. The flux of DM Eq.~(\ref{DMcoma}) is bounded by
these two fluxes and exclusion curve in the space ($m_s$, $\sin^2\,2\theta$)
is bounded by the two straight lines $m_s^5 \sin^2(2\theta) = \const$ (dashed
lines on Fig.~\ref{fig:coma-center-total-flux}), where $const$ is defined (via
Eq.(\ref{DMcoma})) by the fluxes $F_{1\kev}$ and $F_{10\kev}$ correspondingly.

%%%%%%%%%%%%%%%%%%%%%%%%%%%%%%%%%%%%%%%%%%%%%%%%%%%%%%%%%%%%%%%%%%%%%%
\subsubsection{Coma periphery}
\label{sec:coma-periphery}
\begin{figure}
  \centering
  \includegraphics[width=\linewidth]{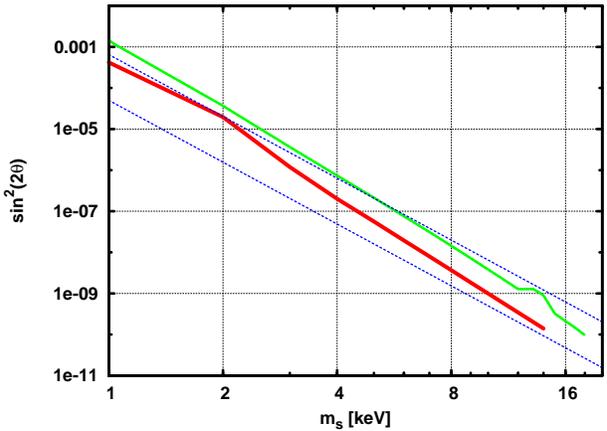} %
  \caption{Exclusion plot based on the restrictions derived from a total flux from Coma 3 peripheral  observation (red solid line) as
    compared with Coma center (green line, identical to the red curve on
    Fig.~\ref{fig:coma-center-total-flux}). Thinner blue lines represent
    $m_s^5\sin^2(2\theta) =\const$ with normalization corresponding to fluxes
    at 1~keV and 7~keV.}
  \label{fig:coma3-total-flux}
\end{figure}

\begin{figure}
  \centering
  \includegraphics[width=\linewidth]{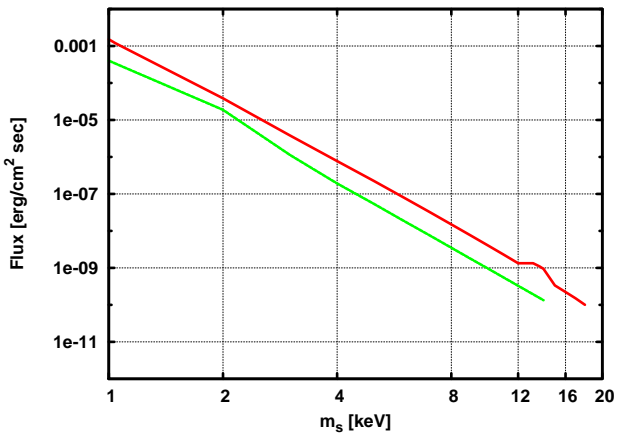} %
  \caption{Comparison between Coma center (red curve) and Coma 3 (green curve)
    fluxes (per 200 eV energy bin).}
  \label{fig:coma-center-vs-coma3-flux}
\end{figure}

%%%%%%%%%%%%%%%%%%%%%%%%%%%%%%%%%%%%%%%%%%%%%%%%%%%%%%%%%%%%%%%%%%%%%%%
As discussed in Section~\ref{sec:theory}, to improve the bound on total flux,
we move away from a center of the cluster.  Unfortunately, modern X-ray
telescopes are normally ``narrow FoV'' instruments and, e.g. the XMM pointing
toward the center of Coma cluster covers only the region of the size $r\sim
r_0$.  Therefore we use one of the observations from Coma mosaic (May-June
2000)~\citep{Briel:00,Neumann:02}.  Although the flux from Coma 3 periphery
falls by a factor 20, as compared to the Coma center\footnote{For Coma 3
  observation $F_{1\kev} = 2\times 10^{-12}$~erg/cm$^2\cdot\mathrm{s}$,
  $F_{10\kev} \sim 10^{-13}$ ~erg/cm$^2\cdot\mathrm{s}$.}, we see that relative change
between 1~keV and 10~keV is again an order of magnitude (c.f.
Fig.~\ref{fig:coma-center-vs-coma3-flux}).  For Coma 3 observation (distance
between the centers of two observations is $39'$ and FoV radius is $\sim
12'$), the geometry factor in Eq.~(\ref{DMcoma}) is equal to
$g_\mathrm{Coma3}=0.18$ which improve an exclusion plot
Fig.~\ref{fig:coma3-total-flux} by about factor of 5.

%\footnotetext{\textbf{To Andrey: get data from 7 to 10 keV}}

\subsubsection{Analysis for Virgo cluster}
\label{sec:total-m87}

We repeat the analysis of the previous Section for the case of center of Virgo
cluster (M87). For this cluster again the surface brightness is well-fitted by
the $\beta$-model with $\beta \simeq 0.4$ and core radius\footnote{We adopt
  the distance to M87 $D_L = 18$~Mpc ($z=0.0044$)} $r_0\simeq 1.6'=8$~kpc and
$S_0 = 5\times 10^{-12}\,\mathrm{\frac{erg}{cm^2\cdot s\cdot
    arcmin^2}}$~\citep{Fabricant:83,Young:02}. The mass in the center of the
Virgo cluster can be described by the isothermal $\beta$-model (see
e.g.~\cite{Nulsen:95}). Using various measurements of the mass distribution in
the center of the Virgo cluster~\citep{Nulsen:95,Evrard:96,McLaughlin:99}, we
obtain
\begin{equation}
  \label{eq:9}
  \rho_{0,\dm} \simeq 6\times 10^{16}\;\mathrm{\frac{M_\odot}{Mpc^3}}
\end{equation}
and correspondingly flux from the FoV circle with (projected) radius $r$ is
given by
\begin{equation}
\label{DMVirgo}
F_{\dm, \textsc{m}87}\simeq 7.7\times
10^{-9}\,\theta^2\left[\frac{m_s}{1\kev}\right]^5
g\left(\tfrac r{r_0}\right)
\mathrm{\frac{erg}{ cm^2\cdot s}}\;.
\end{equation}
The geometry factor $g(r)$ is defined in~(\ref{eq:11}).

Using the M87 observation by XMM-Newton (ID 0114120101), we extract the total
flux from the central $11'$ of the EPN FoV (centered at M87
galaxy).\footnote{The actual experimental data is shown on
  Fig.~\ref{fig:webspec} in red line} %
The geometry factor for this case is $g_{M87}(11')=6.8$.

\begin{figure}
  \centering
  \includegraphics[width=\linewidth]{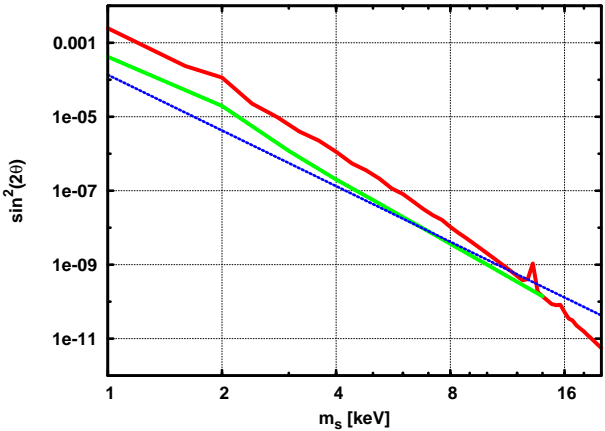} %
  \caption{Restrictions based on the total flux from the circle with the
    radius $11'$ with the center at M87 (red line). Green line represents the
    data for Coma periphery -- same as on
    Fig.~\protect\ref{fig:coma3-total-flux}. Blue line --
    restriction~(\ref{xrbbound}) from~\protect\cite{Boyarsky:05}.} %
  \label{fig:virgo-total-flux}
\end{figure}

Using expression~(\ref{DMVirgo}) for the flux of DM we get a restriction on
parameters of sterile neutrino. The thermal bremsstrahlung of ICM around M87
has temperature about $T=2.54$~keV, which is about 4 times smaller than that
for the Coma cluster. Therefore, the flux drops by about two orders of
magnitude between 1 keV and 10~keV.\footnote{The measured flux at 1~keV is
  $F_{1\,\kev}\simeq 4.8\times 10^{-11}\,\mathrm{erg/cm^2\cdot s}$. This
  number should be compared with the one, obtained from the surface brightness
  profile of~\cite{Fabricant:83}, $F\simeq 4.3\times
  10^{-11}\,\mathrm{erg/cm^2\cdot s}$.}
This leads to an exclusion plot, which is weaker than that from a Coma
periphery at small energies, but gets stronger, than the former restriction at
masses $m_s \gtrsim 12$~keV, as shown on
Fig.~\ref{fig:virgo-total-flux}.\footnote{One should take all total flux
  restrictions at the energies above bremsstrahlung cut-off with the grain of
  salt. Indeed, as will be discussed in more details in
  Section\protect~\ref{sec:chi2-coma}--\ref{sec:chi2-virgo}, the uncertainties
  of measurement of this flux is comparable with the flux itself.}

\subsubsection{ Virgo periphery}
\label{sec:totper-m87}
As we said above, in general periphery of the cluster does not fit into one
XMM FoV.  However, Virgo cluster is a special case in this respect.  Indeed,
in spite of the fact that Virgo cluster is the closest to us galaxy cluster,
the angular size of its core is rather small by the cluster scales. The
surface brightness of the core of the Virgo cluster radius $r_0 =
1.6'$~\cite{Fabricant:83,Young:02} and thus the FoV of the XMM EPN camera
covers about $10 r_0$.  Nevertheless, the ratio of
$\mathcal{S}_{\mathrm{gas}}/\mathcal{S}_\dm$ is changing rather slowly -- as
$(r/r_0)^{-0.58}$
%, $r_0=2$~arcmin$=?$~kpc. This means
%that the XMM FoV which is some 12-15~arcmin in radius covers the
%region of the size $r\sim 6-7r_0$. 
%
%
%From Eq. (\ref{}) one can see that
%at $r=6r_0$ the ratio of DM to continuum surface brightness increases
%by a factor of $6^2=36$ (assuming $\beta=2/3$). 
% For Andrey: beta for Virgo is 0.43

On the other hand, taking ring (with the radii $9'$ and $11'$), rather than
the full circle with the radius $\sim 11'$ leads to the decrease of the
geometric factor in Eq.~(\ref{DMVirgo}) by a factor of about 5, which
compensates the gain almost entirely and therefore, moving away from the
center does not improve the restriction, based on the total flux from the
vicinity of M87.

\subsection{Statistical analysis of the XMM data}
\label{sec:chi2}

In order to further improve the bounds on the sterile neutrino parameters (as
compared to Sections~\ref{sec:total}), we apply the method suggested
in~\cite{Boyarsky:05}. Namely, we use the fact that the spectral shape of the
DM decay line may differ significantly from that of the ICM continuum and
therefore adding a line which is ``too strong'' (e.g. with the flux equal to
that of the continuum) will simply contradict observations.
%Clearly, such a
%method requires some knowledge of the nature of ICM emission spectrum.

In general, typical X-ray spectrum of a galaxy cluster is dominated by the
bremsstrahlung continuum and line emission from the hot intracluster gas.  To
obtain a good fit of observed data to physical model (such as e.g.
MEKAL~\citep{Mewe:86,Liedahl:95} or Raymond-Smith~\citep{Raymond:77}), one
needs to take into account that their parameters (e.g. temperature and
abundances) vary with radius and even with azimuthal angle; that the ICM
plasma is often described by multi-component models; etc. Therefore, often the
sophisticated data analysis is required to extract the ``best-fitting''
parameters for these models and one can only fit these models to the data,
coming from small annular sectors of the FoV.

However, for our purposes one needs \emph{any} ``best-fit'' model to the data,
gathered from previously described annular or circular regions of FoV. We are
not interested in values or distributions of temperatures, abundances, etc. of
the ICM plasma.  Rather, we want to put restrictions on parameters of the
sterile neutrino, based on \emph{non-detection} of the presence of the decay
line against an emission background.  Both from experimental data and from
theoretical models, one can see that X-ray spectrum of galaxies has a form of
continuum plus separate lines. Knowledge of physics (the correct choice of
model) allows to relate various parameters of the model (such as positions and
intensities of the lines, abundances, etc.)  to each other, however, the
structure of the model remains the same: continuum emission plus lines.

Therefore, we introduce a phenomenological model for the cluster spectrum
which provides a good fit to the X-ray data, but at the same time has a
limited number of parameters so that one can easily constrain the maximal
possible contribution of the DM decay line into the X-ray emission from the
cluster.  The restriction is much stronger in those regions of the spectrum,
where one has no lines -- the non-detection of the decay line against
monotonic continuum, essentially puts the restriction on the flux of the line
comparable to the error of the continuum flux.  At the regions, where
emissions lines are present, the restriction of statistical analysis is much
weaker and the flux of DM decay is essentially equal to the excess flux above the continuum.

%{\bf I do not understand how this can be weaker - I would say that it
%should coinside with the bound obtained from the total flux. Misha}

The models and method of their construction are described in
Appendices~\ref{sec:model-coma}--\ref{sec:model-ring}.  These models provide a
satisfactory fit to the observed X-ray spectra of Coma and Virgo clusters.
Obviously, addition of a strong line feature with the integrated line flux
equal to the model flux within a given energy bin would destroy the fit to the
data.  Moreover, it would provide a non-satisfactory fit already at much
smaller DM line flux, of the order of the statistical error of the data in the
narrow energy bin.

%The presence of many X-ray emission lines all over the 1-10~keV energy band in
%the model spectra for the intracluster gas makes it extremely difficult to
%constrain the presence of an additional line contribution of non-plasma
%origin. That is why we prefer in the present work to introduce a
%phenomenological model for the cluster spectrum which provides a good fit to
%the X-ray data, but  at the same time has a very limited number of
%parameters so that one can easily constrain the maximal possible contribution
%of the DM decay line into the X-ray emission from the cluster.\footnote{
%

In order to quantify the above qualitative observation we use the following
algorithm to find the maximal possible contribution of the DM decay line at a
given energy:
\begin{enumerate}
\item We fix the parameters of a
phenomenological model to their best fit values.
\item We introduce an additional model component which is a Gaussian
line at the energy $E_0$ with the width $\Delta E$ equal to the energy
resolution of the instrument ($70-150$~eV for XMM EPN  camera and
15\%$(E/60$~keV$)^{-0.5}$ for Beppo-Sax PDS).
\item Then we increase the normalization of the additional Gaussian until 
  it starts to distort the overall fit to the data. As a simple criterium for
  this we use the increase of reduced $\chi^2$, which leads to the vanishingly small
  \emph{null hypothesis probability}. The actual value of $\Delta\chi^2$
  depends of course on the number of degrees of freedom and has to be chosen
  individually for every data set. We confirm by the visual inspection that
  for such a distortion, fitted model does not follows the data anymore in the
  region of $E\sim E_0$.

%plot (lower panel of
%  Fig.~\ref{fig:spectrum}a) of the fit of the phenomenological model alone
%  never exceeds $\pm 4$. 

%The distortion of the fit by the additional Gaussian
%  is clearly visible on the $\delta\chi$ plot when $\delta\chi(E_0)$ reaches
%  the vlaue of $-5$. We use this as a marginally possible value to put the
%  limit on the normalization of the additional Gaussian.
\item We make a ``scan'' over the line energy $0.5\kev\le E_0\le 10$~keV (and
  $15~\kev \le E_0 \le 40~\kev$ for Beppo-Sax data) to find the maximal
  possible normalization of the Gaussian as a function of energy.
\end{enumerate}
As discussed above and in \cite{Boyarsky:05}, the only assumption about the
empirical model, which we make, is that the power-law regions of the spectrum
are actual monotonic bremsstrahlung and do not contain ``dips''~see
Fig.~\ref{fig:spectrum}b.

In the Sections~\ref{sec:chi2-coma}--~\ref{sec:chi2-virgo} below we compare
this method with that of the Section~\ref{sec:total} for the case of Coma and
Virgo cluster.

%%%%%%%%%%%%%%%%%%%%%%%%%%%%%%%%%%%%%%%%%%%%%%%%%%%%%%%%%%%%%%%%
\subsubsection{Bounds from statistical analysis of the Coma cluster
  observations.} 
\label{sec:chi2-coma}
%%%%%%%%%%%%%%%%%%%%%%%%%%%%%%%%%%%%%%%%%%%%%%%%%%%%%%%%%%%%%

The simple model for Coma center X-ray emission is provided in
Appendix~\ref{sec:model-coma}. 
\begin{figure}
  \centering
  \includegraphics[width=\linewidth]{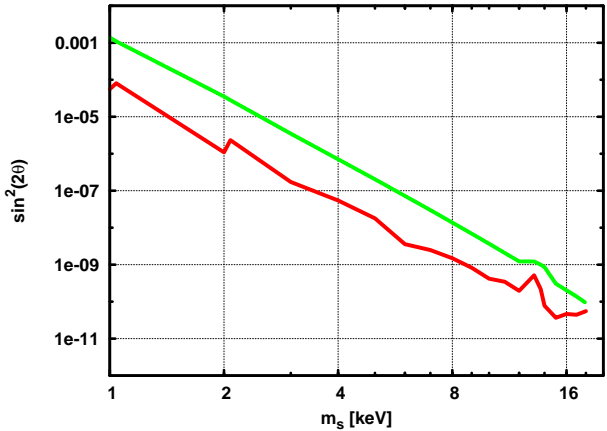} %
  \caption{Red line: exclusion plot based on statistical analysis from Coma center
    observation (circular region of the radius $700''$ around the center of
    the cluster). Green line -- restriction based on the total flux from the
    same region (curve from Fig.~\protect\ref{fig:coma-center-total-flux}).}
  \label{fig:comaC-chi2}
\end{figure}
The results of statistical analysis (and demonstration of its improvement over
the total flux restrictions) are shown on the Fig.~\ref{fig:comaC-chi2}. One
can see that overall one gets about an order of magnitude improvement over the
results of Section~\ref{sec:total-coma}. One can also see ``spikes'' at $E
\simeq 1$~keV, $E\simeq 6.5$~keV, etc. ($m_s \simeq 2$~keV,$13$~keV,etc.) --
positions of the strong emission lines in the Coma spectrum (see
Appendix~\ref{sec:model-coma} for their identification). This is expected.
Adding a line in the energy region where data is monotonic and well described
by the smooth power-law bremsstrahlung emission distorts the fit much more
than adding the line with the same strength to the region of multiple element
emission lines.

\begin{figure}
  \centering
  \includegraphics[width=\linewidth]{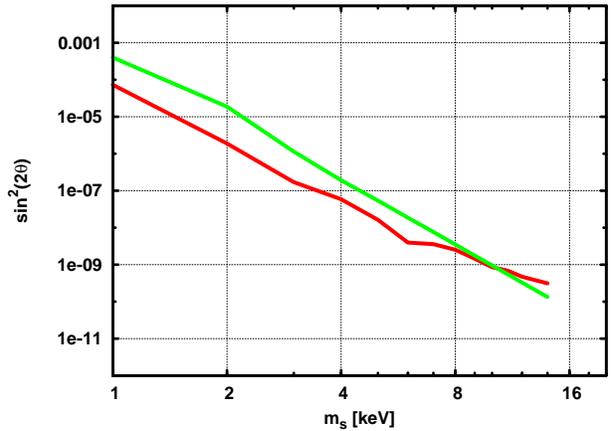} %
  \caption{Red line: exclusion plot based on statistical analysis from Coma 3
    peripheral observation of Section~\protect\ref{sec:coma-periphery}. Green
    line -- restriction based on the total flux from the same region (curve
    from Fig.~\protect\ref{fig:coma3-total-flux}).}
  \label{fig:coma3-chi2}
\end{figure}
Similarly, the results of statistical analysis of the data from Coma 3 region
are shown on Fig.~\ref{fig:coma3-chi2}. Here, the improvement is not so
drastic as in case of Fig.~\ref{fig:comaC-chi2}. The reason for that is
obvious -- the statistics for Coma 3 is lower and therefore the fitted curve
can be distorted stronger, without spoiling reduced $\chi^2$ too much. From
looking at the Fig.~\ref{fig:coma3-chi2} one can see that the proposed
statistical analysis is in some sense complimentary to that of the total flux
analysis: namely, while it is stronger at small masses ($m_s \lesssim 10$~keV,
it gets more and more weak as $m_s$ increases -- tendency which is quite
opposite to that of the total flux restriction. The reason for that is clear
-- at large energies there is almost no X-ray emission from cluster, and the
difference between a very weak signal from X-ray emission and background has
very large uncertainty. Therefore, total flux restriction for $m_s \gtrsim 10
\kev$ is provided only for illustrative purposes and the results of the
statistical analysis should be considered as a correct limit.\footnote{We have
  used the Beppo-Sax observation of Coma cluster, to obtain constraints in the
  range $15\kev \le E \le 40$~keV. The obtained restrictions (both statistical
  and ``total flux'') are about an order of magnitude weaker than restrictions
  from XRB.}

%%%%%%%%%%%%%%%%%%%%%%%%%%%%%%%%%%%%%%%%%%%%%%%%%%%%%%%%%%%%%%%%
\subsubsection{Bounds from statistical analysis of the 
  of Virgo cluster.}
\label{sec:chi2-virgo}
%%%%%%%%%%%%%%%%%%%%%%%%%%%%%%%%%%%%%%%%%%%%%%%%%%%%%%%%%%%%%

\begin{figure}
  \centering
  \includegraphics[width=\linewidth]{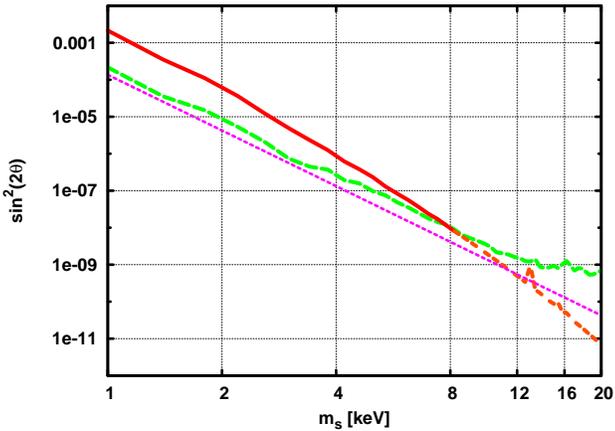} %
  \caption{Restrictions from the statistical analysis of the periphery (ring
    $9'$--$11'$) around the M87 (green dashed line). This restriction is about
    an order of magnitude better that the one, coming from the total flux,
    measured from the same region (red solid line) at small energies ($m_s
    \lesssim 8$~keV and then become worse (red dashed line), as statistical
    error of the data grows. This is due to the fact that for $E\gtrsim 4\kev$
    total flux is measured with uncertainty comparable or exceeding the flux
    itself.  The straight line represent the XRB
    restrictions~(\ref{xrbbound}).}
  \label{fig:virgo-chi2}
\end{figure}
Let us apply the same strategy to the XMM pointing toward M87/Virgo cluster.
For this we consider the spectrum collected from a thin ring $9'<r<11'$ around
the M87. We attempt to add a DM decay line to the resulting spectrum,
described in Appendix~\ref{sec:model-ring} and find the maximal strength of
the line that does not deform the spectral fit (in the sense described in the
previous sections). The resulting exclusion plot is shown by the dashed lines
in Fig.~\ref{fig:virgo-chi2}.  We see that the statistical analysis improves
the restrictions over that of the total flux for small masses (i.e. for which
$E=m_s/2 \lesssim T$, where $T\sim 2.5$~keV is the temperature of the ICM for
the center of the Virgo cluster, while gets significantly worse for higher
energies. As discussed in Section~\ref{sec:chi2-coma}, this is due to the fact
that the errors in flux are comparable (or even exceed) the flux itself and
therefore the results of the statistical analysis should be considered as the
correct limit.

One can also see that the restriction from the Virgo is not as strong as the
one, which we obtained from the Coma center (shown on
Fig.~\ref{fig:comaC-chi2}). This is due to the same reason as above: Coma
cluster has higher temperature and the data, collected from a large field, has
bigger statistical significance, hence smaller errors.

\section{Results}
\label{sec:results}

\begin{figure}
  \centering %
  \includegraphics[width=\linewidth]{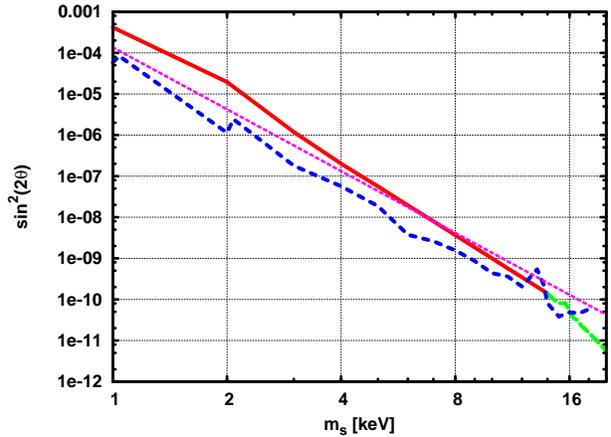} %
  \caption{Exclusion plot for parameters of sterile neutrino, based on
    analysis of the cluster of the Coma and Virgo clusters of galaxies. The
    red solid line represents the restriction, coming from the Coma periphery
    (``Coma 3'') observation, while the green dashed line, coming from
    observation of Virgo cluster, provides stronger restriction for $m_s
    \gtrsim 14\kev$. The blue dashed curve represents the result of
    statistical analysis of the Coma center. The straight magenta line
    represents the XRB restrictions from~\protect\cite{Boyarsky:05}.}
  \label{fig:result}
\end{figure}

In this Section we summarize the restrictions on parameters of sterile
neutrino, obtained from analysis of the Virgo and Coma cluster of galaxies.
Based on analysis of the core and peripheral regions of these clusters we
obtain the following results:
\begin{enumerate}
\item The most robust restriction is based on the requirement that the flux of
  DM did not exceed the total X-ray flux in the given energy bin (chosen to be
  $\Delta E = 200$~eV in our case). The best restriction comes from the
  observation of the Coma periphery (XMM-Newton observation ``Coma 3'' (ID
  0124710301) with EPIC EPN camera). It provides an exclusion plot, shown in
  the solid red line on the Fig.~\ref{fig:result}.
  
\item For $m_s \gtrsim 14$~keV, the restriction from the Virgo cluster becomes
  stronger (due to the fact, that the temperature of the intracluster gas in
  the Virgo cluster is several times lower, than that of the Coma. This
  results in an improvement, shown in the green dashed line, continuing
  previous bound above $m_s \simeq 14$~keV on Fig.~\ref{fig:result}.
  
\item By analyzing the form of the spectrum, one can refine the restrictions
  on parameters of sterile neutrino (as discussed in details in
  Section~\ref{sec:chi2}). From all the data used in this work, the best
  result was obtained in analysis of the Coma center region (see
  Section~\ref{sec:chi2-coma} and Appendix~\ref{sec:model-coma}). The
  resulting exclusion region is shown as dashed blue line on the
  Fig.~\ref{fig:result}. It provides an improvement by a factor 2--4 over the
  XRB result of \cite{Boyarsky:05} in the region $2\kev \lesssim m_s\lesssim
  10\kev$.  One can qualitatively understand why Coma center observations
  gives the best result for statistical analysis. As discussed in the
  Section~\ref{sec:chi2-coma}, the more monotonic is the data, the better the
  method of statistical analysis works. The result significantly improves over
  the total flux restriction only in the regions where ``there are no lines''.
  Due to the fact that Coma temperature is 4 times bigger than that of Virgo,
  the phenomenological model for the Coma cluster requires only three
  additional lines (as discussed in details in Appendix~\ref{sec:model-coma}).
\end{enumerate}

%%%%%%%%%%%%%%%%%%%%%%%%%%%%%%%%%%%%%%%%%%%%%%%%%%%%%%%%%%%%%%%%%%%%%%%%
%\begin{figure}
%  \includegraphics[width=\columnwidth,angle=0]{exclusion_SAX}
%  \caption{Bounds on the sterile neutrino parameters obtained from the
%  analysis of Coma (solid curves)  and Virgo clusters(dashed curves). 
%  Green curves: bound from the requirement
%  that the flux in the DM decay line does not exceed the flux in the
%  continuum emission at the line energy. Red curves: bound obtained
%  from the statistical analysis described in Section 4. For comparison
%also shown the bound obtained from the analysis of the X-ray
%background data (black).}
%  \label{fig:exclusion}
%\end{figure}
%%%%%%%%%%%%%%%%%%%%%%%%%%%%%%%%%%%%%%%%%%%%%%%%%%%%%%%%%%%%%%%%%%%%%%%%
 
%%%%%%%%%%%%%%%%%%%%%%%%%%%%%%%%%%%%%%%%%%%%%%%%%%%%%%%%%%%%%
\section{Discussion.}
\label{sec:concl}
%%%%%%%%%%%%%%%%%%%%%%%%%%%%%%%%%%%%%%%%%%%%%%%%%%%%%%%%%%%%%

In this work we analyzed the restrictions on the parameters of the sterile
neutrino, as a dark matter candidate, coming from the observations of several
clusters of galaxies. Below we compare our results with those of other authors
and discuss what improvements on restriction plots one expects to obtain from
galaxy clusters data.

\begin{figure}
  \centering
  \includegraphics[width=0.8\linewidth,angle=-90]{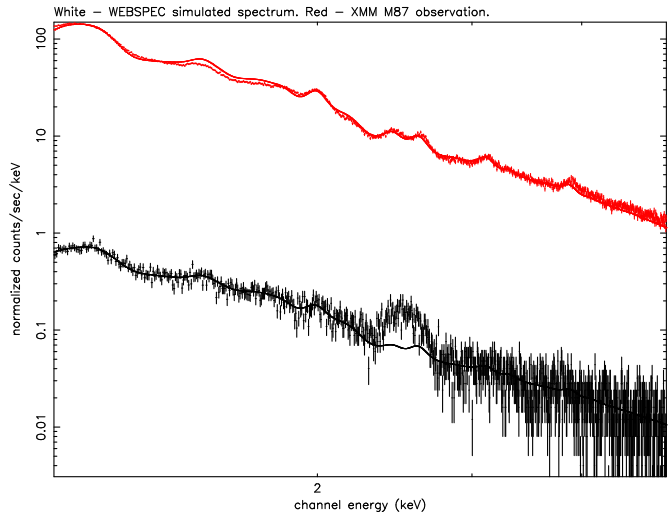} %
  \caption{Comparison of WEBSPEC simulations of~\protect\cite{Abazajian:01b}
    (lower black line) with actual XMM observation of M87 -- red line.} %
  \label{fig:webspec}
\end{figure}

\begin{enumerate}
\item We see that this work brings improvement by a factor 2--4 (in the region
  $2\kev\lesssim m_s \lesssim 10\kev$) over the previous analysis
  of~\cite{Boyarsky:05} (see Fig.~\ref{fig:result}). This should be compared
  with the results of~\cite{Abazajian:01b,Abazajian:05a}, where stronger
  constraints, derived from the restriction of the total flux of the Virgo
  cluster, were claimed. Below, we explain the reason for this difference.
  
  Assuming $\theta \simeq 6\times 10^{-5}$, \cite{Abazajian:01b} argued that
  if the sterile neutrino mass would be $\gtrsim 5$~keV, the ``enormous'' DM
  decay line observed in the core of Virgo galaxy cluster (or, more precisely
  from its cD galaxy M87) would dominate over the continuum flux. To
  illustrate this,~\citet{Abazajian:01b} modeled (via WEBSPEC interface) the
  emission of the M87 by MEKAL model with the temperature $T=2.5$~keV and
  overall normalization of the flux in the 2-10~keV interval equal to
  $1.5\times 10^{-12}$erg/(cm$^2\cdot\mathrm{s}$), the number taken
  from~\cite{Boehringer:01}. This value for the emission flux from the center
  of Virgo is, however, two orders of magnitude \emph{below} the actual flux,
  as one can see from the literature~\citep{Fabricant:83,Arnaud:98} or by
  comparing these numbers with the actual experimental data (e.g.  observation
  of M87 with XMM in 2001 (obs.  ID 0114120101). From the latter one can
  measure that the flux in the range 2-10~keV is some two orders of magnitude
  higher ($\sim 10^{-10}$~erg/cm$^2\cdot\mathrm{s}$). While the discussed in
  the paper of~\citet{Abazajian:01b} flux of DM is taken from the whole FoV of
  Chandra (which has an area $\sim 300~\mathrm{arcmin}^2$), the number
  $1.5\times 10^{-12}$~erg/cm$^2\cdot\mathrm{s}$ in the \citet{Boehringer:01}
  refers to the emission from the active nucleus of the central galaxy of
  Virgo cluster, M87, collected from the central 10 arcsec of the XMM (not
  Chandra) field of view. For the mixing $\theta \simeq 6\times 10^{-5}$, used
  in \cite{Abazajian:01b}, the flux of DM is about two orders of magnitude
  below the MEKAL flux in the energy bin around $E=m_s/2=2.5$~keV (see
  Fig.~\ref{fig:webspec}). In the later work \cite{Abazajian:05a}, the same
  data was used to provide an exclusion plot in the $(m_s,\,\sin^2 2\theta)$
  parameter space of the form $\sin^2(2\theta) \lesssim 1.0\times 10^{-5}
  (\kev/m_s)^2$.

\item Statistical analysis of the Coma center and Coma 3 data, gave very close
  restrictions. However, Coma 3 represents only one of the observations of
  Coma mosaic, which cover the annular region between $\sim 10'$ and $40'$
  around the center of the Coma cluster. By combining these results, one
  effectively increases observation time by about a factor of 10. This should
  lead to the modest improvement of the statistical restriction by about
  $\sqrt{10}$. We leave this for the future work.
  
\item The restrictions, based on the total flux, can be improved if one moves
  to the periphery of the cluster. The farther away from the center one moves,
  the stronger becomes ``signal-to-background'' ratio (especially in the case
  of the Coma cluster, c.f. Eq.~(\ref{sb})). Therefore, one should try to move
  away from the center of the cluster. The maximal distance where one can move
  i restricted by the requirement that the DM decay line is still stronger,
  than the XRB background (otherwise one does not gain anything, compared to
  the restriction~(\ref{xrbbound}) from XRB).
  
\item At the energies above 15 keV, other missions (such as INTEGRAL) should
  be used to obtain the restrictions on parameters of sterile neutrino.
\end{enumerate}

\subsection*{Acknowledgements}

We would like to thank K.~Abazajian and I.~Tkachev for useful comments.  The
work of M.S. was supported in part by the Swiss Science Foundation. The work
of O.R. was supported in part by European Research Training Network contract
005104 "ForcesUniverse" and by a \emph{Marie Curie International Fellowship}
within the $6^\mathrm{th}$ European Community Framework Programme.

\appendix

%%%%%%%%%%%%%%%%%%%%%%%%%%%%%%%%%%%%%%%%%%%%%%%%%%%%%%%%%%%%%%%%
\section{Simple model for the X-ray spectrum of the core of Coma cluster.}
\label{sec:model-coma}
%%%%%%%%%%%%%%%%%%%%%%%%%%%%%%%%%%%%%%%%%%%%%%%%%%%%%%%%%%%%%

%%%%%%%%%%%%%%%%%%%%%%%%%%%%%%%%%%%%%%%%%%%%%%%%%%%%%%%%%%%%%%%%%%%%%%%%
\begin{figure*}
  \includegraphics[width=\linewidth,angle=0]{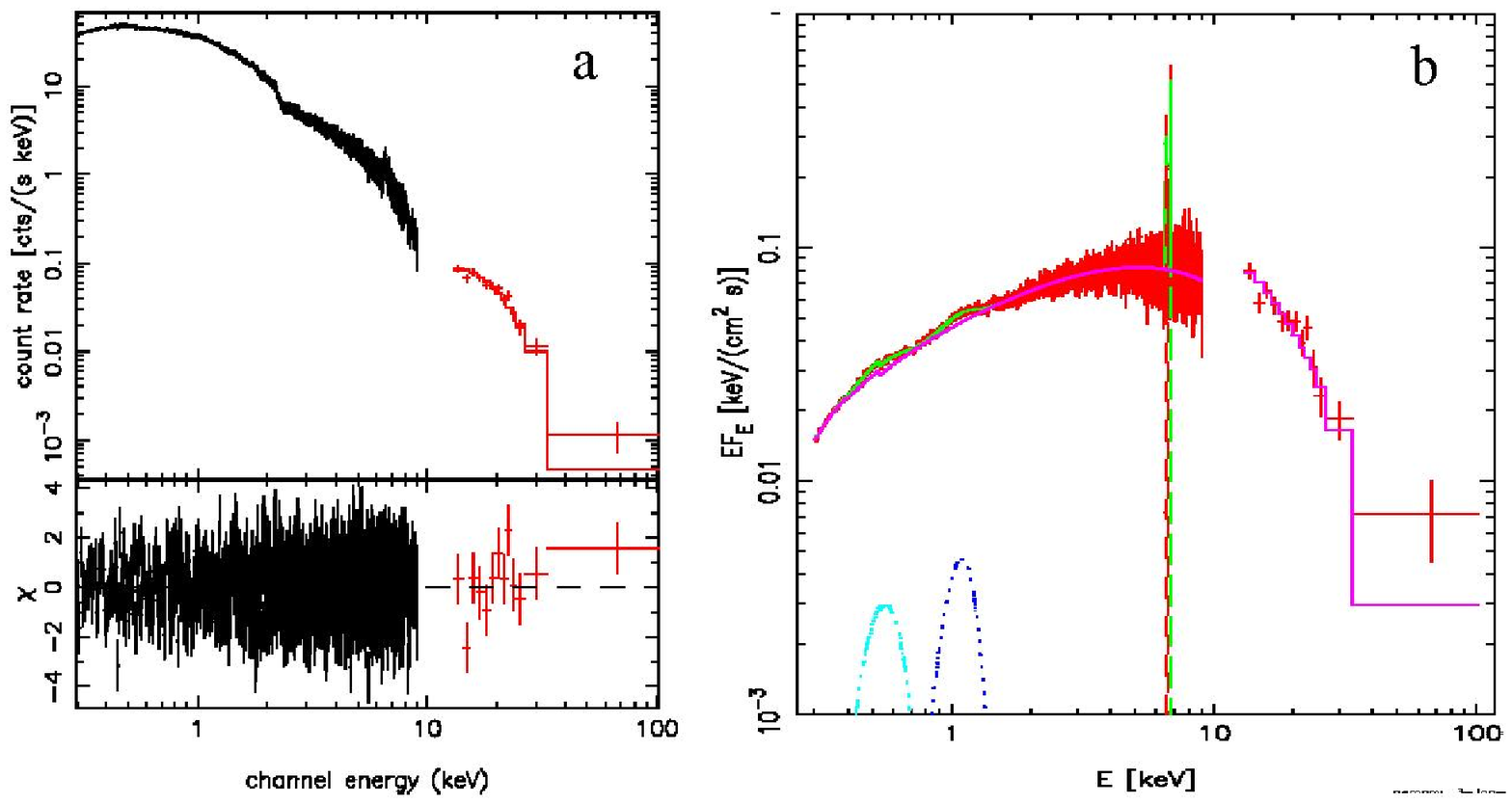}
  \caption{Folded (a) and unfolded (b) XMM-Newton and Beppo-SAX spectrum of
    the Coma galaxy cluster. The simple phenomenological model used to fit the
    data consists of a cutoff powerlaw modified by the galactic absorption and
    four Gaussian lines. The model parameter values are given in the text.}
  \label{fig:spectrum}
\end{figure*}
%%%%%%%%%%%%%%%%%%%%%%%%%%%%%%%%%%%%%%%%%%%%%%%%%%%%%%%%%%%%%%%%%%%%%%%%

%\textbf{\textbf{\ldots} We analyze the XMM-Newton observation of the core of
%  Coma cluster, observational ID 0124711401, June 2000 and Beppo-SAX
%  observation \textbf{???}}

The simple phenomenological fit to the XMM-Newton PN and Beppo-SAX PDS data is
shown in Fig. \ref{fig:spectrum}. The model consists of a powerlaw with a
photon index $\Gamma=1.36\pm0.04$ modified at high energies by an exponential
cutoff with the cut-off energy $E_{cut}=8.3\pm 0.7$~keV and by the absorption
with column density $N_H=(1.5\pm0.3)\times 10^{20}$~cm$^{-2}$ (which is
roughly the galactic absorption in the direction of Coma cluster). Several
bright X-ray lines are visible in addition to the cut-off powerlaw component.
They are the red-shifted 6.7~keV and 7.0 keV Fe K$\alpha$ and K$\beta$ lines
as well as 0.5~keV and 1.0~keV O and Ne lines. All the components of the model
are shown together with the unfolded 0.3-50~keV spectrum of the central part
of the Coma cluster in the right panel of Fig.  \ref{fig:spectrum}. The
reduced $\chi^2$ of the fit is $\chi^2=1.04$.

%%%%%%%%%%%%%%%%%%%%%%%%%%%%%%%%%%%%%%%%%%%%%%%%%%%%%%%%%%%%%%%%
\section{Simple model for the X-ray spectrum of the core of Virgo cluster.}
\label{sec:model-m87}
%%%%%%%%%%%%%%%%%%%%%%%%%%%%%%%%%%%%%%%%%%%%%%%%%%%%%%%%%%%%%

Modeling the spectrum of the central part of Virgo cluster turns out to be
more complicated because more lines are evident in the 1-5~keV part of the
spectrum. This can be explained by lower, compared to Coma, temeperature of
Virgo cluster, $T\simeq 2-3$~keV. Besides, Virgo cluster is a ``cD galaxy
dominated'' cluster, which means that the diffuse X-ray emission from the
cluster is strongly dominated by emission from the central giant elliptical
galaxy M87. The size of the X-ray core of M87 is much smaller than the size of
X-ray core of Coma cluster, so that in spite of the fact that Virgo is about 5
times closer to the Earth than Coma, the angular size of its core, $r_0\simeq
2$~arcmin, is 5 times less than the one for Coma. This, in turn, implies that
the temperature of the intracluster (or intragalactic) medium varies
significantly over the XMM FoV, which extends up to $6-7r_0$. Finally, Virgo
cluster is not a relaxed galaxy cluster, and assumptions about hydrostatic
equilibrium, which one has to evoque to compute the DM distribution may not
hold.

The above facts make the modelling of the X-ray spectrum collected from the
entire XMM FoV difficult. However, we achieve a satisfactory phenomenological
fit the the Virgo cluster spectrum with the model similar to the one used in
the previous Section for Coma cluster analysis. It consists of the powerlaw
with the photon index $\Gamma=1.28$ modified at high energies by an
exponential cut-off at $E_0=2.22$~keV and at low energies by photoelectric
absorption on the Galactic hydrogen column density $N_H=2.5\times
10^{20}\,\mathrm{cm}^{-2}$.  Several lines have to be added to the fit, to
make it acceptable. In particular, Fe K$\alpha$ and K$\beta$ lines of Fe (6.7
and 7.9~keV), line of O at 0.64~keV, a set of lines at 1~keV from the cooling
flow in the center of Virgo cluster, etc. (7 lines altogether).

%%%%%%%%%%%%%%%%%%%%%%%%%%%%%%%%%%%%%%%%%%%%%%%%%%%%%%%%%%%%%%%%
\section{Simple model for the X-ray spectrum of the ring with the radius
  $9'-11'$ around M87.}
\label{sec:model-ring}
%%%%%%%%%%%%%%%%%%%%%%%%%%%%%%%%%%%%%%%%%%%%%%%%%%%%%%%%%%%%%

We repeat the procedure already used for the analysis of the central parts of
Coma and Virgo clusters for the present case. Namely, we find a simple
phenomenological fit to the broad-band spectrum which in this case is given by
the cutoff powerlaw with $\Gamma=1.29$ and $E_0=2.24$, modified by the
Galactic absorption. Again, we add several lines (as in the case of the
previous Appendix) to make fit acceptable.

\end{document}

\end{document}